# Interactions between basal dislocations and $\beta'_1$ precipitates in Mg–4Zn alloy: mechanisms and strengthening

R. Alizadeh[1] and J. LLorca[1,2]

[1] IMDEA Materials Institute, c/Eric Kandel 2, 28906 Getafe, Madrid, Spain.
[2] Department of Materials Science, Polytechnic University of Madrid. E.T.S. de Ingenieros de Caminos, 28040 Madrid, Spain.

**Abstract**

The mechanisms of dislocation/precipitate interaction as well as the critical resolved shear stress were determined as a function of temperature in a Mg–4 wt.% Zn alloy by means of micropillar compression tests. It was found that the mechanical properties were independent of the micropillar size when the cross-section was > 3 × 3 µm². Transmission electron microscopy showed that deformation involved a mixture of dislocation bowing around the precipitates and precipitate shearing. The initial yield strength was compatible with the predictions of the Orowan model for dislocation bowing around the precipitates. Nevertheless, precipitate shearing was dominant afterwards, leading to the formation of slip bands in which the rod precipitates were transformed into globular particles, limiting the strain hardening. The importance of precipitate shearing increased with temperature and was responsible for the reduction in the mechanical properties of the alloy from 23 ºC to 100 ºC.

1. **Introduction**

Low density, good castability, high specific strength and stiffness, and reasonable cost make Mg alloys attractive for different applications. Despite these advantages, poor formability and ductility at room-temperature as well as the limited yield strength restrict their use as a conventional engineering material for structural applications [1–5]. The origin of these limitations in Mg and Mg alloys is related to their hexagonal close-packed (hcp) crystal structure, which leads to large differences in the critical resolved shear stresses (CRSS) to activate different deformation modes at low temperature (below 100 °C). The CRSS for <*a*> basal slip in pure Mg is ≈ 0.5 MPa at room-temperature [6–8], much lower than the CRSS for <*a*> prismatic slip (≈ 18 MPa) and for <*c+a*> pyramidal slip (≈ 40 MPa) [9,10]. As a result, basal slip is always the dominant deformation mechanism in Mg alloys, limiting the strength and also the ductility because of the localization of deformation in clusters of grains suitably oriented for basal slip [11]. Moreover, basal slip cannot account for the deformation along the *c* axis of the Mg lattice which is normally accommodated by twinning across the $\{10\bar{1}2\}$ planes. Twining is a polar mechanism which is only activated by stress states leading to an extension of the *c* axis and, thus, gives rise to a strong plastic anisotropy in textured Mg alloys that also reduces the formability and the ductility.

Obviously, the strength and ductility of Mg alloys can be improved by increasing the CRSS of basal slip through either solid solution and/or precipitate hardening. Nevertheless, accurate experimental data on the effect of solute atoms or precipitates on the CRSS for basal slip are scarce because they are difficult to obtain from mechanical tests in polycrystalline samples because of the superposition of different strengthening contributions (grain boundaries, latent hardening, texture) that cannot be easily isolated [12]. The most reliable data were obtained from mechanical tests on single crystals strengthened by solid solution [13–15] or precipitates [14–16] that can be tested along specific crystallographic directions to ensure that only basal slip is activated. Nevertheless, this technique is extremely time consuming and expensive because of the cost associated with the manufacturing of the single crystals.

Alternatively, micromechanical testing techniques based on compression of single crystal micropillars or nanopillars manufactured by focused ion beam milling from polycrystals have been applied in the last decade to explore the different deformation modes of Mg [17–22] and Mg alloys [23–27]. More recently, micropillar compression tests have been used to determine the influence of solid solution elements on the CRSS for basal slip in Mg alloys [28]. The main limitation of these micromechanical testing techniques to obtain reliable values of the CRSS is found in the strong size effects that appear when the volume of the specimen tested is of the order of tenths of $\mu m^3$ [20,29–32]. The origin of this size effect has been thoroughly analyzed in the past [30] and comes about as a result of the interaction between the critical dimensions of the specimen (i.e. the diameter of the micropillar) and material length scale that controls the strength (i.e. the average distance between dislocations or precipitates). Size effects appear when the latter approaches the former and tend to be very strong in well-annealed fcc and hcp metals and alloys deformed along the soft modes. However, size effects during micropillar compression tests are negligible in precipitation hardened alloys when the precipitate spacing is much lower than the micropillar dimensions [33].

In this investigation, the effect of precipitates on the CRSS for basal slip at 23 ºC and 100 ºC was analyzed by means of micropillar compression tests in an Mg–4 wt.% Zn alloy which was aged at different temperatures to produce different precipitate distributions. This alloy was selected because the $\beta'_1$ precipitates, which form as elongated rods along the $c$ axis of the Mg matrix, lead to one of the strongest age hardening responses among Mg alloys [34,35]. Although the mechanical properties of the Mg-Zn alloy are well-established [36], the CRSS for basal dislocations in this system has only been studied in [24,37]. Chung and Byrne [37] measured the CRSS in Mg-5.1 wt.% Zn alloy from -269ºC to 27ºC using tensile tests of single crystals in alloys aged at 200ºC for 4 and 28 h. The presence of $\beta'_1$ precipitates was ascertained by transmission electron microscopy but no information was provided regarding the precipitates (diameter and spacing) and the dislocation/precipitate interactions. Wang and Stanford [24] measured the CRSS by compression of circular micropillars of 2 μm in diameter at 23ºC of a Mg- 5 wt.% alloy aged at 150ºC during 8 days. They reported the actual values of the precipitate diameter and spacing but they did not check the effect of the micropillar diameter on the CRSS. They showed one low magnification transmission electron micrograph (Fig. 8b [24]), which was compatible with precipitate shearing, but the details of the dislocation/precipitate interaction mechanisms were not studied. In this paper, the effect of the micropillar size on the mechanical response was carefully analyzed to obtain results of the CRSS that were independent of the micropillar size and could be compared with the precipitation hardening models available in the literature. Moreover, the details of the dislocation/precipitate interactions were carefully examined as a function of temperature and provided new insights in the limited efficiency of precipitate strengthening in Mg alloys, as compared with other metallic alloys.

## 2. Experimental methods

*2.1. Processing*

A Mg–4 wt.% Zn alloy was prepared from high purity Mg (99.90 wt.%) and Zn (99.99 wt.%) pellets. They were melted using a graphite crucible in an induction furnace (VSG 002 DS, PVA TePla) under a protective Ar atmosphere to avoid oxidation. The melt was held at 750 °C for about 10 min to provide a homogeneous composition and was poured into a copper die, installed inside the furnace chamber, using a tilt-casting system to minimize casting defects and the melt turbulence. The cast rods, with diameter of 12 mm and height of 150 mm, were put in quartz capsules under an Ar protective atmosphere and homogenized at 450 °C for 15 days to homogenize the composition and to promote grain growth. The samples were cut into small cylinders of 12 mm in diameter and 18 mm in height, and hot compressed at 400 °C up to 30 % strain in a Gleeble 3800 thermo-mechanical simulator. Deformation was applied to increase the dislocation density in order to promote grain growth in the following annealing step. Afterwards, they were subjected to a solution heat treatment at 450 °C for 20 days in quartz capsules under an Ar protective atmosphere. Finally, they were aged at 149 °C for 100 h and at 204 °C for 16 h to create precipitates with different size. These aging conditions were chosen from the literature [38] and hardness experiments.

*2.2. Microstructural Characterization*

Cross-sections of the aged samples were prepared for electron backscatter diffraction (EBSD) analysis. They were polished using a MD-Mol cloth with a 3 μm diamond paste, followed by polishing with MD-Nap cloth with 0.25 μm diamond paste. Finally, the samples were chemically polished using a solution of 75 ml ethylene glycol, 24 ml of distilled water and 1 ml of nitric acid to remove any residual surface damage and also to reveal slightly the grain boundaries. EBSD measurements were carried out in a dual-beam field emission gun scanning electron microscope (SEM) (Helios Nanolab 600i FEI) equipped with an Oxford-HKL electron back scattered system. They were conducted at an accelerating voltage of 20 kV and a beam current of 2.7 nA, using a step size of about 2-4 μm, depending on the magnification. Afterwards, the grain size and crystal orientation obtained from the EBSD data were used to choose the most appropriate grains for studying basal dislocations/precipitate interactions.

The morphology of the precipitates was studied both before and after deformation through transmission electron microscopy (TEM) (FEI Talos F200X) at 200 kV. Lamellae with an approximate thickness of 100 nm were milled with a focused ion beam (FIB) (Helios Nanolab 600i FEI) using a Ga+ ion beam with an accelerating voltage of 30 kV.

*2.3. Micro-mechanical characterization*

Micropillars were milled with a focused ion beam in grains suitably oriented for basal slip when deformed in compression. To this end, grains with Schmid factor (SF) higher than 0.45 for the basal slip and below 0.25 for twinning and prismatic slip were selected. The micropillars were milled with a square cross-section to facilitate the observation and analysis of the slip traces on the lateral surfaces. Different pillar sizes with cross sections of 1×1, 3×3, 5×5 and 7×7 $\mu m^2$ were prepared to investigate the size effect on the deformation. Hereafter, for simplicity, these pillars will be denominated 1, 3, 5 and 7 µm micropillars, respectively. In all cases, the height-to-length aspect ratio was approximately between 2:1 and 3:1 to avoid buckling for higher aspect ratios, or non-uniform stresses along the length for lower aspect ratios [39,40]. The FIB milling process was performed using the annular technique with a Ga+ ion beam at an accelerating voltage of 30 kV. A beam current of 9.3 nA was used for the coarse milling step, and a small ion beam current of 40 pA was employed for the final polishing step to minimize the ion implantation on the surface. During the final milling step, the stage was tilted 2° more with respect to the ion beam axis to reduce the taper angle, which was always below 1º. It should be noted that annular milling (as compared with lathe milling) minimizes the influence of Ga+ implantation on the deformation mechanisms and mechanical properties of the micropillars [41]. Moreover, no obvious effect of Ga+ implantation was detected near the surface of the pillars.

The micropillar compression tests were performed at both room-temperature (23 °C) and 100 °C using a TI950 TriboindenterTM (Hysitron, Minneapolis, MN). The load was applied with a diamond flat punch tip with diameter of 10 µm installed in the low-load transducer. All tests were performed in the displacement-control mode at a strain rate of ≈$10^{-3}$ $s^{-1}$ up to a maximum strain of 10%. At least two tests were performed for each micropillar size and temperature to ensure the reproducibility of the results. The experimental load-displacement curves were corrected to account for the extra compliance due to the elastic deflection of the pillar into the substrate following the Sneddon method [29,42]. The corrected curves were converted to the engineering stress ($\sigma$)-engineering stain ($\varepsilon$) curves based on the initial cross-section area and height of the pillars measured after the milling process. They were transformed into resolved shear stress ($\tau_{RSS}$)-strain ($\varepsilon$) curves ($\tau_{RSS} = \sigma \times SF$) to compare the results obtained in different grains.

After deformation, the micropillars were carefully examined using high-resolution scanning electron microscopy (HR-SEM) to analyze the slip traces that appeared on the lateral surfaces. Additionally, several of the deformed micro-pillars were further analyzed by transmission Kikuchi diffraction (TKD) to elucidate the mechanisms involved during deformation. The lamellae used for TKD were prepared using the same methodology used for milling lamellae for TEM studies. The TKD mapping was carried out in the FEI Helios NanoLab 600i SEM with the specimen oriented horizontally. The electron beam was operated at 15 kV with a beam current of 2.7 nA. The TKD maps were collected with step size of 100 nm.

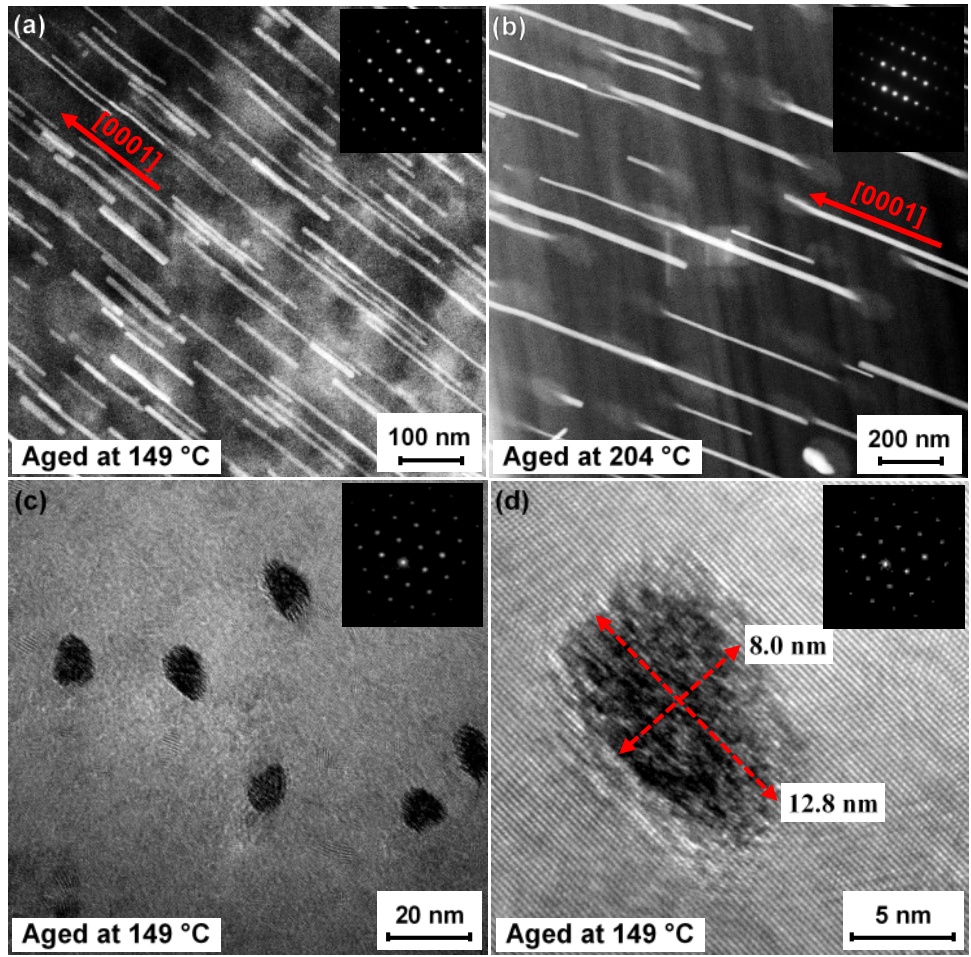

**Fig. 1.** STEM micrographs showing the elongated precipitates parallel to the *c*-axis of the Mg matrix in: (a) alloy aged at 149 °C for 100 h and (b) aged at 204 °C for 16 h. The red arrows in (a) and (b) are parallel to the *c*-axis of the Mg matrix. (c,d) HRTEM micrographs showing the shape and size of the precipitates in the material aged at 149 °C for 100 h in a section parallel to the basal plane of the Mg matrix. The corresponding selected area electron diffraction patterns of the Mg matrix are included in each figure.

## 3. Results
### 3.1. Morphology of the precipitates

High magnification TEM micrographs of the material after aging at 149 °C for 100 h and at 204 °C for 16 h are presented in Fig. 1, showing the general morphology of the precipitates. The scanning transmission electron microscopy (STEM) tomography movie of the precipitates in the material aged at 149 °C for 100 h is also shown in the supplementary material (Fig. S1). The rod-shape precipitates grew along the *c*-axis of matrix (Figs. 1a, 1b, and S1), and presented a more or less equiaxed cross section in the basal plane (Figs. 1c and 1d). The length and diameter distributions of the precipitates were determined based on the STEM micrographs of lamellas parallel and perpendicular, respectively, to the *c* axis of the Mg matrix using ImageJ and are shown in Fig. 2. The diameter was determined from the total area of the cross-section assuming a circular shape. The average length and diameter of the precipitates in the material aged at 149 °C for 100 h were 146 ± 10 nm and 9.7 ± 2 nm, respectively. These results are very close to those reported by Wang and Stanford [24] in a Mg–5 wt.% Zn subjected to similar aging conditions. Coarser precipitates were found in the alloy aged at 204 °C for 24 h, with average length and diameters of 235 ± 10 nm and 15.0 ± 2 nm, respectively. The average aspect ratio of the precipitates was similar for both aging conditions.

The volume fraction of the precipitates was determined from the area fraction of the precipitates (measured with ImageJ) in the STEM micrographs of lamellae perpendicular to the *c* axis of the Mg matrix because the lamella thickness (around 100 nm) was equivalent to the average precipitate length. Thus, the volume fractions of precipitates were estimated as 2.6 ± 0.6 % and 1.8 ± 0.4 % for the materials aged at 149 °C and 204 °C, respectively. Assuming that the precipitates were distributed in a triangular pattern on the Mg basal plane, the average precipitate spacing in the plane perpendicular to the *c* axis of the Mg matrix, $\lambda$, can be calculated as [43]

$$\lambda = \sqrt{\frac{\pi}{\sqrt{3}f}}\, d \qquad (1)$$

where *f* is the volume fraction of the precipitates and *d* the average diameter which can be found above for both aging conditions. Thus, the average precipitate spacing was 82 ± 10 nm and 152 ± 17 nm for the alloys aged at 149 ºC and 204 ºC, respectively.

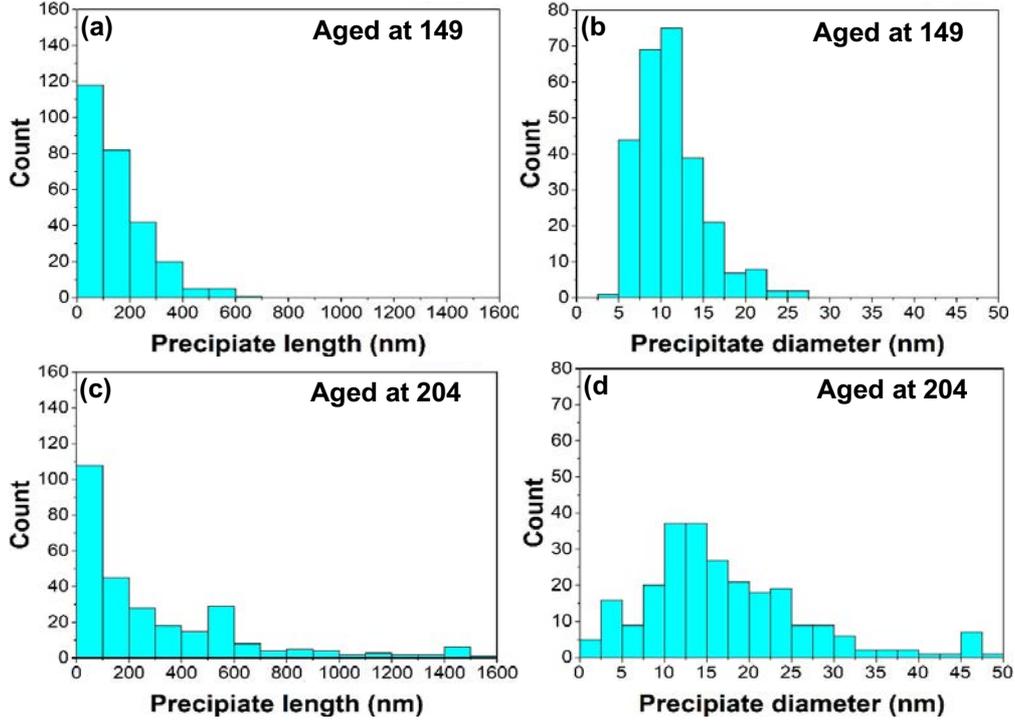

**Fig. 2.** (a) Precipitate length and (b) diameter distribution after aging at 149 °C for 100 h. (c) Precipitate length and (d) diameter distribution after aging at 204 °C for 16 h.

The types of the precipitates in Mg-Zn alloys have been analyzed in the past [34,37,38]. $\beta_1'$ precipitates appear as long rod-shaped precipitates parallel to the *c* axis of the Mg matrix, while $\beta_2'$ precipitates are short rod-shaped precipitates parallel to basal plane of the Mg matrix and $\beta$ precipitates are block-shaped particles with random orientations. The morphology and orientation of the precipitates in Fig. 1 was only compatible with $\beta_1'$ precipitates. The diffraction patterns of the precipitates obtained by TEM were not free from the matrix contribution above and beneath the precipitate but the high-angle annular dark-field imaging-STEM analysis indicated that $[2\bar{1}\bar{1}0]_{\beta_1'} \parallel [0001]_{Mg}$ and $[0001]_{\beta_1'} \parallel [2\bar{1}\bar{1}0]_{Mg}$, as expected from $\beta_1'$ precipitates. This Laves phase has a hcp crystal structure with a chemical composition of MgZn$_2$. The growth habit plane is given by $(0001)_{\beta_1'} \parallel (11\bar{2}0)_{Mg}$ with a coincident direction $[11\bar{2}0]_{\beta_1'} \parallel [0001]_{Mg}$ [38]. The *a*-axis of the $\beta_1'$ precipitate grows parallel to the *c* axis of the magnesium matrix and the *c* axis of $\beta_1'$ is parallel to one of the three equivalent $[11\bar{2}0]$ directions in the matrix. Moreover, the compact basal plane of Mg is parallel to the prismatic plane of the precipitate, $(0001)_{Mg} \parallel (2\bar{1}\bar{1}0)_{\beta_1'}$ [34]. Since the spacing along the *a* axis of $\beta_1'$ closely matches the spacing along the *c*-axis of the alloy matrix (i.e., good coherency) the $\beta_1'$/Mg interface parallel to the *c*-axis of the matrix is a well ordered dislocation boundary [38].

The $\beta_1'$ phase is very stable and breakdown of coherency occurs only after long aging times, leading to the apparition of the equilibrium $Mg_2Zn_3$ phase with a random orientation in the matrix [37]. No evidence of the equilibrium $Mg_2Zn_3$ phase was found in the samples aged at both temperatures. In addition, scattered precipitates of the $\beta_2'$ phase ($MgZn_2$), which grows with a habit plane $(0001)_{\beta_2'} \| (0001)_{Mg}$ [37], were occasionally found but their volume fraction was negligible in comparison with $\beta_1'$.

### 3.2. Effect of micropillar size on the mechanical properties

The mechanical properties of the aged alloys were measured by means of micropillar compression tests at 23 ºC and 100 ºC. In general, the reproducibility of the tests was excellent. Two representative $\tau_{RSS}$-$\varepsilon$ curves corresponding to 3 μm micropillars tested at 23 ºC and two 5 μm micropillars tested at 100 ºC of the alloy aged at 149 ºC are plotted in Figs. 3a and 3b, respectively. Up to five tests were carried out in selected cases when the results of the two first tests were not very close (perhaps because of bad alignment between the flat punch and the micropillar surface) to obtain accurate estimations of the CRSS. Moreover, the strategy to determine the initial CRSS from the $\tau_{RSS}$-$\varepsilon$ curves is indicated in Fig. 3a. A straight line was fitted to the linear region of the elastic part of the $\tau_{RSS}$-$\varepsilon$ curve (marked with A) starting where the concave shape of the curve (due to inhomogeneous contact between the flat punch and the micropillar) ends. The CRSS was given by line B, where $\tau_{RSS}$-$\varepsilon$ curve deviates from the straight line A.

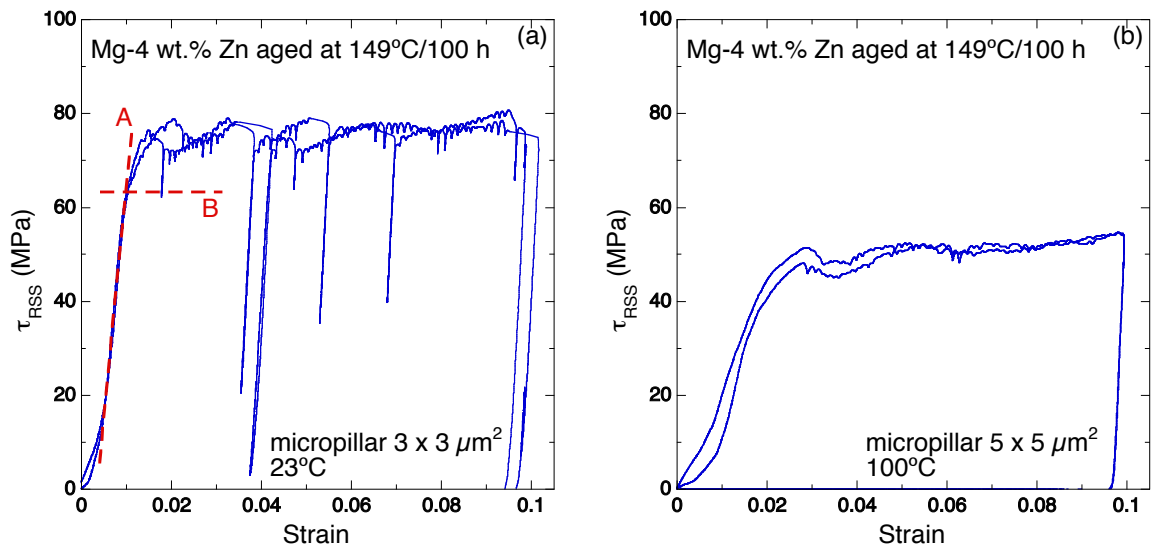

**Fig. 3.** Reproducibility of the stress-strain curves of the alloy aged at 149 °C. (a) Compression of 3 μm micropillars at 23 ºC. (b) Compression of 5 μm micropillars at 100 ºC. The deviation from linearity criterion (as depicted in Fig. 3a) was used to calculate the initial CRSS.

The effect of the micropillar size on the $\tau_{RSS}$-$\varepsilon$ curves was analyzed in the alloy aged at 149 ºC and the corresponding curves are plotted in Fig. 4 for micropillars with dimensions in the range of 1 to 7 μm deformed at 23 ºC. The objective of the size effect analysis was just to determine the minimum micropillar cross-section to avoid large size effects and, thus, to obtain the bulk properties of the alloy. The results in Fig. 4 show that micropillars of 5 x 5 μm² provided an optimum compromise because the size effects were very small (the difference with the 7 x7 μm² micropillars was only of 5 MPa in the initial CRSS and negligible in the CRSS at 4% strain). Moreover, the curves of the 1 μm micropillar presented large strain bursts, which were associated with the localization of the deformation in one dominant slip band. The higher flow stress and the strain bursts are due to the lack of mobile dislocations to carry the plastic deformation because of the small pillar size. Thus, elastic stresses are built up to nucleate more dislocations and plastic deformation is localized in a single plane when one dislocation source becomes active. The $\tau_{RSS}$-$\varepsilon$ curves of the micropillars of 3 to 7 μm showed smaller strain bursts and the plastic deformation was initially distributed along the pillar, indicating a homogeneous deformation. Strains bursts were negligible for the 5 μm and 7 μm pillars (Figs. 3b and 4). For all micropillar sizes, plastic deformation took place along the basal plane and no traces of the activation of twinning or slip in other systems were detected. Thus, the mechanical properties obtained with micropillars of 5 μm (which is between 30 to 60 times longer than the precipitate spacing) were free from artifacts due to size effects and could be used to determine the CRSS of the Mg-Zn alloys aged at different temperatures.

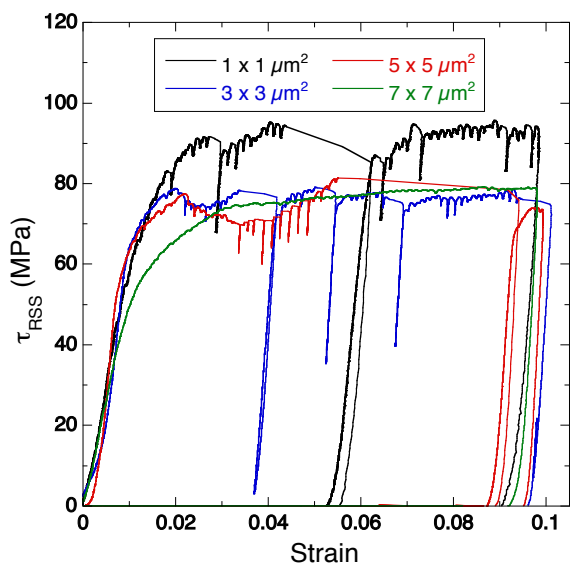

**Fig. 4.** Influence of the micropillar dimensions on the resolved shear stress-strain curves of the alloy aged at 149 ºC and tested at 23 ºC.

### 3.3. Mechanical properties at 23 ºC and 100 ºC

The mechanical properties of the Mg–4Zn alloy aged at 149 ºC and 204 ºC were determined by means of compression tests in micropillars of 5 μm at 23 ºC and 100 ºC. The corresponding $\tau_{\text{RSS}}$-$\varepsilon$ curves are shown in Figs. 5a and 5b and the initial CRSS as well as the CRSS at 4% strain are summarized in Table 1. The features of the curves were very similar regardless of the aging condition and temperature. There was an initial elastic region followed by plastic yielding. The shear stress necessary to produce basal slip increased dramatically as a result of the presence of the $\beta_1'$. The CRSS of basal slip for the same alloy when the Zn was in solid solution was measured using the same methodology in [28] and it was found that the contribution of solid solution hardening to the CRSS was ≈ 4 MPa. Strain hardening was observed in all cases after the yield point up to an applied strain of ≈ 2%. Afterwards, the flow stress remained constant up to the maximum applied strain of 10% and this behavior was associated to the localization of deformation along one of several slip bands across the micropillar.

The highest values of the initial CRSS and of the CRSS at 4% at 23 ºC and 100 ºC were measured in the alloy aged at 149 ºC, which contained a higher volume fraction of smaller precipitates. Moreover, a large reduction in the mechanical properties was observed for both alloys with temperature from 23 ºC to 100 ºC. The initial CRSS and the flow stress at 4% strain decreased with temperature by 34% and 35%, respectively, in the alloy aged at 149ºC, and by 38% and 23% in the alloy aged at 204ºC. These reductions are very important, as precipitate strengthening is normally considered to be independent of the temperature in so far the precipitates are stable. It should be noted that the experimental values of the initial CRSS values of the material aged at 204 °C tested at 100 ºC presented a large scatter, very likely due to inhomogeneous contact between the flat punch and the micropillar surface at the beginning of the test. However, both curves were very close to each other after initial yielding.

To the authors' knowledge, there are only two papers that report the CRSS for basal slip in precipitation-hardened Mg-Zn alloys and this information was included in Table 1 for comparison. Chun and Byrne [37] measured the CRSS in Mg-5.1 wt.% Zn alloy using tensile tests of single crystals in alloys aged at 200 ºC for 4 and 28 h. The presence of $\beta_1'$ precipitates was ascertained by TEM but not detailed quantitative analysis of the precipitate diameter and spacing was provided. Wang and Stanford [24] measured the CRSS by compression of circular micropillars of 2 μm in diameter of a Mg- 5 wt.% alloy aged at 150 ºC during 8 days. They reported the actual values of the precipitate diameter and spacing but they did not check the effect of the micropillar diameter on the CRSS. So, the experimental information in our paper provides -for the first time- reliable values of the CRSS and of the precipitate dimensions and spacing that can be used to check the validity of different strengthening models.

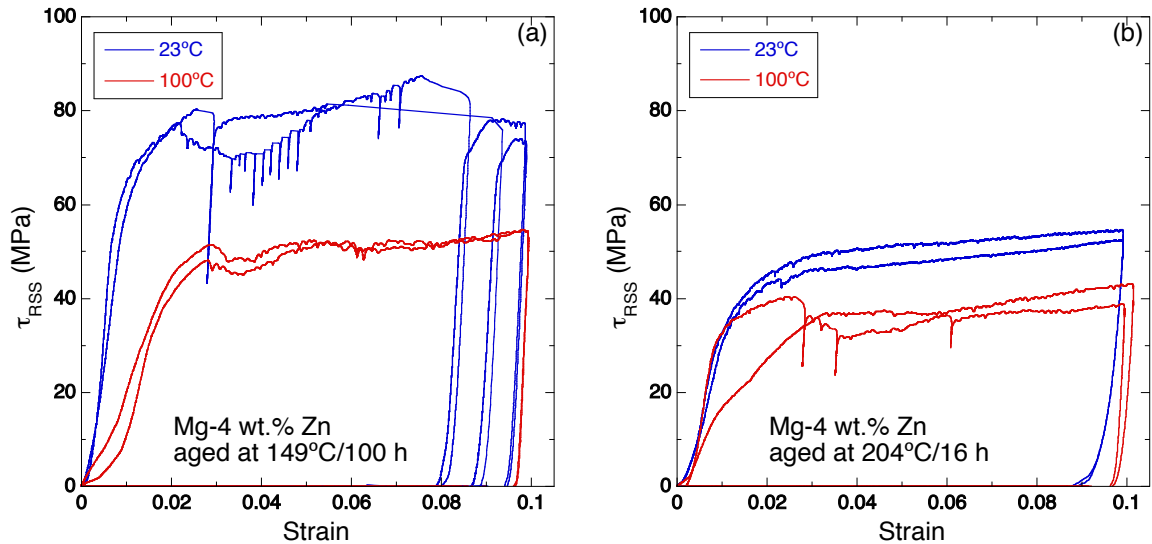

**Fig. 5.** Resolved shear stress vs. strain curves of micropillars of 5 μm compressed at 23 ºC and 100 ºC. (a) Mg–4 wt.% Zn alloy aged at 149ºC for 100 h. (b) Mg–4 wt.% Zn alloy aged at 204ºC for 16 h.

Table 1 – Initial CRSS and CRSS at 4% strain for basal slip of the Mg–Zn alloys as a function of the heat treatment and test temperature. CRSS for basal slip on similar alloys reported in the literature are also included for comparison.

| Composition (wt.%) | Aging condition | Test method | Test temperature (°C) | Initial CRSS (MPa) | CRSS at 4% strain (MPa) | Reference |
|---|---|---|---|---|---|---|
| Mg-4Zn | 100 h at 149 °C | Micro-pillar compression (5×5 μm$^2$) | 23 | 55.1 | 75.6 | Present work |
| Mg-4Zn | 100 h at 149 °C | Micro-pillar compression (5×5 μm$^2$) | 100 | 36.5 | 48.8 | Present work |
| Mg-4Zn | 16 h at 204 °C | Micro-pillar compression (5×5 μm$^2$) | 23 | 35.6 | 48.6 | Present work |
| Mg-4Zn | 16 h at 204 °C | Micro-pillar compression (5×5 μm$^2$) | 100 | 22.1 | 37.6 | Present work |
| Mg-5Zn | 192 h at 150 °C | Micro-pillar compression (diameter of 2 μm) | 23 | 47.7 | - | [24] |
| Mg-5.1Zn | 4.4 h at 200 °C | Tensile deformation of single crystal | 27 | 10 | - | [37] |
| Mg-5.1Zn | 28 h at 200 °C | Tensile deformation of single crystal | 27 | 18 | - | [37] |

## 3.4. Deformation micromechanisms

The deformation micromechanisms were analyzed by means of TEM and TKD in the deformed micropillars. To this end, lamellae parallel to the vertical surfaces of the micropillars were milled by FIB from the middle of the pillars. Bright-field TEM micrographs of the micropillar aged at 149 ºC after deformation at 23 ºC are shown in Figs. 6a and 6b. Several slip bands parallel to the basal plane (marked with yellow broken lines) can be observed in the higher magnification micrograph (Fig. 6b). The deformation was more localized between bands (1) and (2), and it was accompanied by formation of a visible step at the micropillar surface (Fig. 6c). The TKD map depicted in Fig. 6d, together with the selected area electron diffraction patterns of points (d1) and (d2) in Fig. 6b, demonstrate that deformation was dominated by basal slip. Changes in crystal orientation due to the activation of twinning were not found even in the severely deformed area between bands (1) and (2).

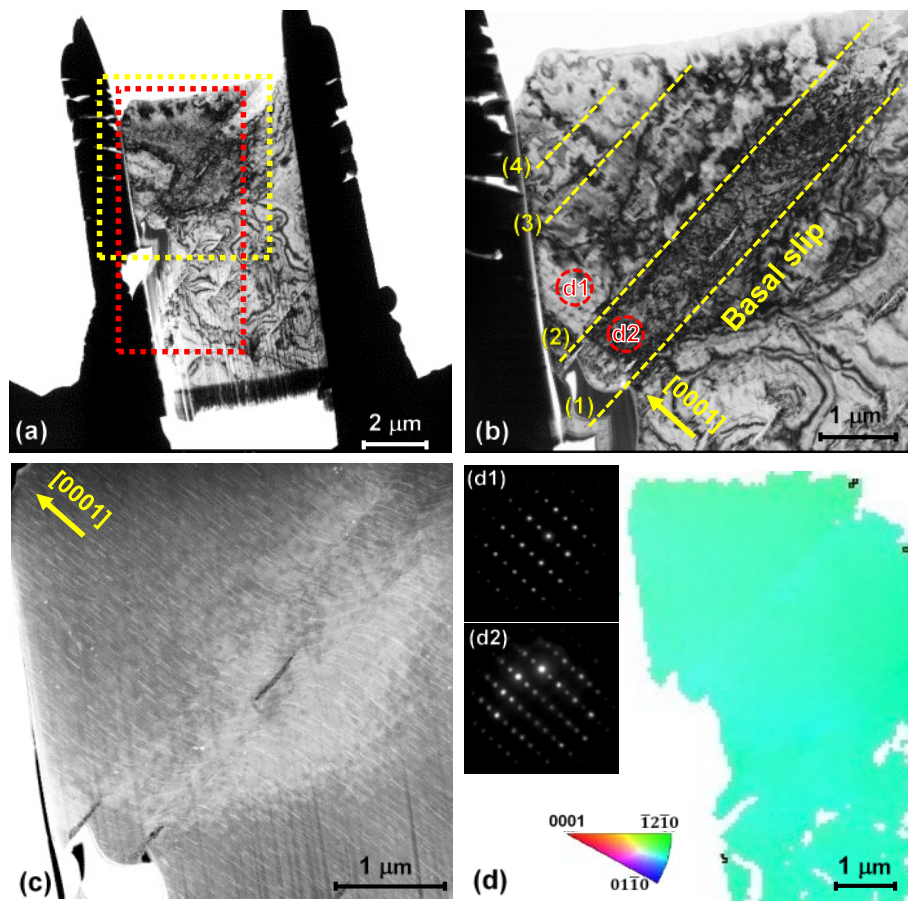

**Fig. 6.** Analysis of the thin lamella extracted from the 5 µm micropillar of the material aged at 149 °C after deformation at 23 ºC. (a,b) Low magnification bright field TEM micrographs (c) STEM micrograph. (d) TKD map including the diffraction patterns corresponding to (d1) and (d2) areas in (b). The regions surrounded by yellow and red dashed lines in (a) are shown in (b) and (d), respectively.

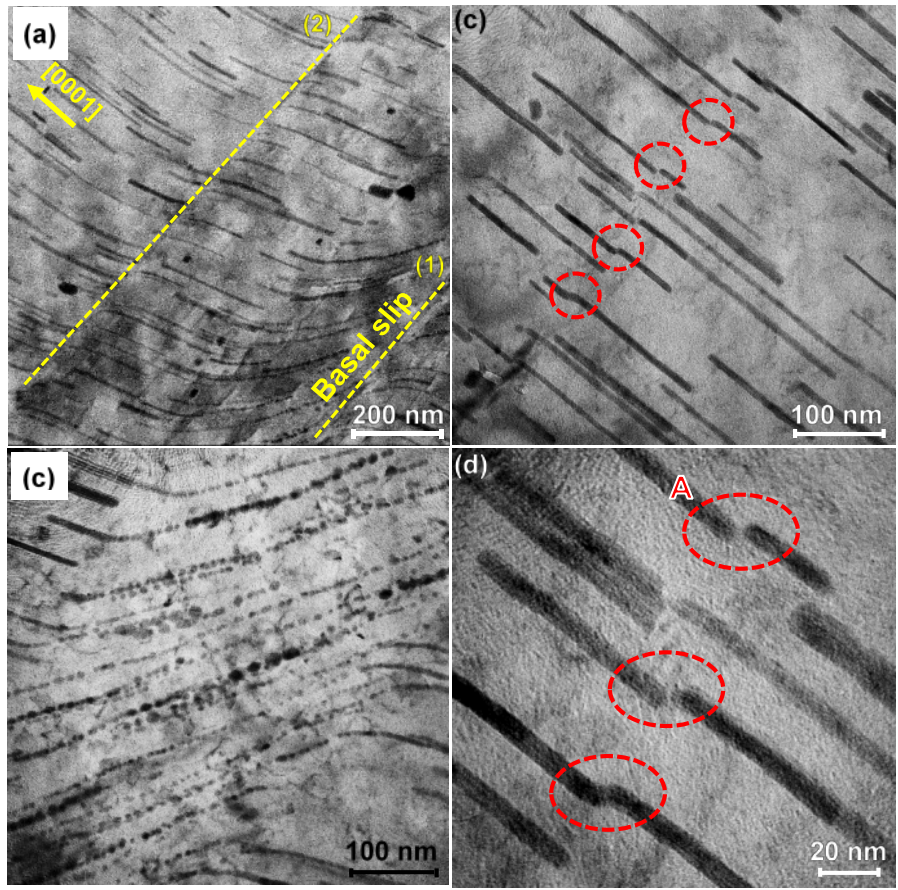

**Fig. 7.** Bright field TEM micrographs showing the dislocation/precipitate interactions in the alloy aged at 149 ºC and deformed at 23 ºC. (a) Severe deformation of the precipitates between slip lines 1 and 2 in Fig. 6b. (b) Precipitate shearing around slip line 3 in Fig. 6b. (c) Morphology of the fragmented precipitates in the severely deformed region. (d) High magnification image showing the mechanisms of precipitate shearing by basal dislocations.

The mechanisms of dislocation/precipitate interaction can be ascertained in the higher magnification TEM micrographs presented in Fig. 7. The $\beta_1'$ precipitates that intercepted the slip bands 1 and 2 (Figs. 7a) were sheared by the basal dislocations. This mechanism was also observed also in regions that were not subjected to very severe deformation (Fig. 7b). Moreover, some precipitates were cut into multiple small fragments in the region between the slip bands 1 and 2 as a result of precipitate shearing by basal dislocations (Fig. 7c). The mechanisms of precipitate shearing by basal dislocations are more clearly observed in the higher magnification micrograph presented in Fig. 7d, which shows the formation of a kink in the precipitate. Thus, precipitate shearing does not seem to take place along well-defined crystallographic planes but involves the deformation of the precipitate along a region encompassing several atomic planes. The step measured at the A precipitate in Fig. 7d was 6.2 nm, much longer than the Burgers vector of basal dislocations (0.32 nm). This result indicates that the precipitates were sheared by many dislocations. Moreover, fragmentation of the precipitates is more evident in the STEM tomography video given in supplementary material (Fig. S2). Other precipitates located near the slip band 1 were severely deformed to accommodate the large shear deformation around the slip band but they were not broken, while the $\beta_1'$ precipitates located far from the basal slip bands retained their original shape and orientation and did not participate in plastic deformation (Figs. 7b, 7d and S2).

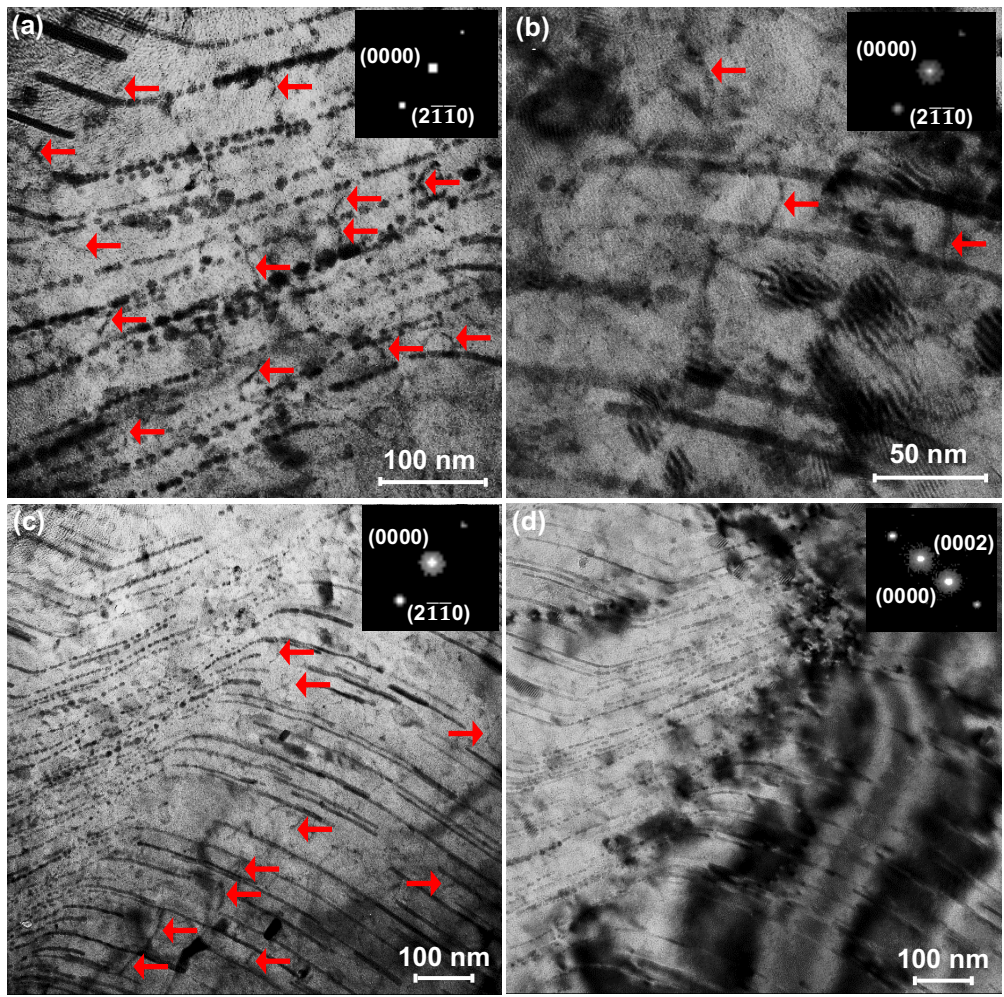

**Fig. 8.** High magnification bright field TEM micrographs of the 5 μm pillar of the alloy aged at 149 °C after deformation at 23 ºC. The small arrows point to basal dislocations pinned by the precipitates. The lamella was parallel to $(01\bar{1}0)$ with g: $[\bar{2}110]$ in (a), (b) and (c), and to $(\bar{2}110)$ with g: $[0002]$ in (d). Basal <a> dislocations can be observed in (a), (b) and (c) (because of the beam condition) but not in (d) because the beam condition is not appropriate to observe <a> dislocations. These results demonstrate that the dislocations (indicated by red arrows) are <a> basal dislocations.

In addition to precipitate shearing, dislocation bowing between the precipitates was also observed in some TEM micrographs (Fig. 8), indicating that precipitate shearing was not the only mechanism of dislocation/precipitate interaction. Basal <a> dislocations could be observed in Fig. 8 according to the two beam condition because the lamella plane was parallel to (01$\bar{1}$0) while g was [$\bar{2}$110]. The dislocation contrast disappeared when the lamella was tilted to the (2$\bar{1}\bar{1}$0) plane and g was parallel to [0002], indicating that they were <a> dislocations. It should be noted that bowing of dislocations between $\beta'_1$ precipitates was also reported at -196ºC in [38].

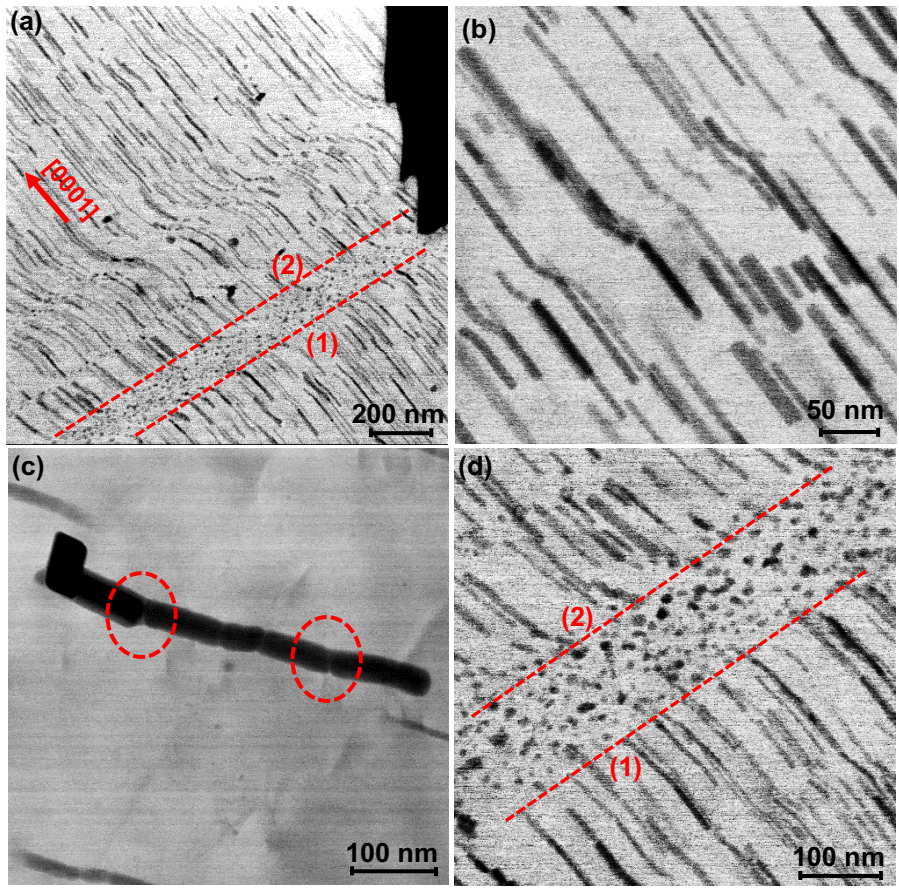

**Fig. 9.** High magnification bright field TEM micrographs of the 5 µm pillar of the material aged at 149 °C after deformation at 100 ºC. (a) Evidence of multiple slips bands, (b) and (c) shearing of $\beta'_1$ precipitates by basal dislocations and (d) precipitates transformed into globular particles within intense slip bands.

The deformation mechanisms in the same alloy after deformation at 100 ºC can be ascertained from the bright field TEM micrographs of the 5 µm pillar in Fig. 9. More slip bands were formed along the micropillar (Fig. 9a) in comparison with micropillars deformed at 23 ºC (Fig. 6b). The deformation was, thus, more homogeneous at 100 ºC and this is in agreement with the smoother stress-strain curves in Fig. 5a. On the contrary, the deformation in the micropillars deformed at 23 ºC was localized in fewer, more intense slip bands, which led to the appearance of stress drops in the stress-strain curves (Fig. 5). Shearing of the precipitates by basal dislocations was also observed at 100 ºC and it was also associated the formation of kinks (Fig. 9b). It is interesting to notice that even very thick precipitates (as the one shown in Fig. 9c with a diameter ≈ 38 ± 3 nm) were sheared by dislocations at different sections along the length. Moreover, precipitates were transformed in globular particles around the more intense slip bands (Fig. 9d). This globular morphology is also evident in two STEM tomography movies presented in the supplementary material (Figs. S3 and S4).

It should be finally noticed that precipitate shearing was not the only interaction mechanism between dislocations and precipitates at 100 ºC. Bowing of basal dislocations around the precipitates was also observed after high-temperature deformation, as shown in Fig. 10 in the material aged at 204 °C. Similar results were also found for the material aged at 149 °C.

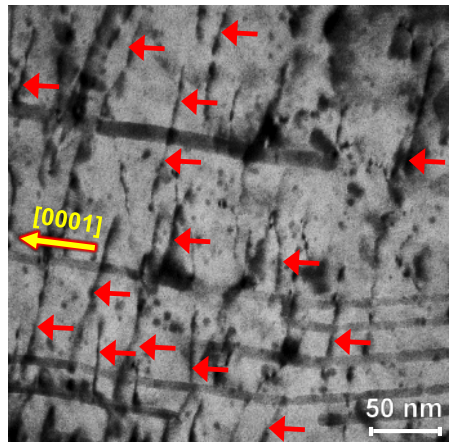

**Fig. 10.** Bright field TEM micrograph of the 5 μm micropillar of the alloy aged at 204 °C after deformation at 100 ºC. Basal dislocations bowing around precipitates are marked with arrows.

### 4. Discussion

#### 4.1. Interaction mechanisms of basal dislocations with $\beta_1'$ precipitates

The mechanical behavior of a Mg-5.1 wt.% Zn alloy aged at 200 ºC was studied as a function of test temperature (from 4.2 to 300 K) by Chun and Byrne [37] by means of mechanical tests of single crystals whose orientations were closer to the center of the inverse pole figure. A significant reduction in the CRSS with temperature was found for the peak-aged and overaged conditions. Moreover, the effect of the strain rate on the CRSS in the peak-aged material was negligible at 4.2 K and increased rapidly with temperature. It should be noted that the microstructure of the alloy in this paper was very similar to our Mg-4Zn alloy aged at 204 ºC for 16 h. These observations, in addition to the low work-hardening rate, led them to conclude that precipitate shearing was the main deformation mechanism, although no physical evidence by TEM was provided. They supported this conclusion from the analysis of the orientation relationship between the $\beta_1'$ precipitates and the Mg matrix because the plane $(2\bar{1}\bar{1}0)$ of the $\beta_1'$ precipitates which has the highest atomic density was parallel to the basal plane of the Mg matrix, facilitating the shearing of precipitates.

Further analysis of the mechanisms of dislocation/precipitate interactions in Mg alloys were based on the Orowan mechanisms [35,43], but evidence of precipitate shearing by dislocations in a Mg-5 wt.% Zn alloy subjected to similar aging conditions was provided by Wang and Stanford [24] after compression of one micropillar of 2 µm. Nevertheless, the small size of the micropillar lead to large size effects (noticed in significant strain bursts) and to very large values of the CRSS. In parallel, recent experimental observations have also reported precipitate shearing by basal dislocations in Mg-Nd [44], Mg-RE [44], Mg-Mn-Nd and Mg-Al [45] alloys. Nevertheless, it should be noticed that the in situ TEM observations of Huang et al. [46] in an Mg-RE alloy showed that basal dislocations have to pile-up at precipitate/matrix interface before they were able to shear the precipitate. These results are in agreement with recent molecular dynamics simulations by Esteban-Manzanares et al. [47] which showed that the first dislocation entered the precipitate but was not able to progress further, leading to the formation of an Orowan loop within the precipitate. The arrival of new dislocations and the stress concentration associated with the dislocation pile-up finally led to the collapse of the Orowan loop and to the shearing of the precipitate.

The experimental observations presented in the previous section provide a clear picture of the interaction of basal dislocations with precipitates in Mg-Zn alloy that can be easily extended to other Mg alloys. Basal dislocations are easily activated due to the low CRSS for basal slip in Mg and they rapidly form loops around the precipitates. Experimental evidence [48] showed that sometimes the dislocations were attracted by the precipitate/matrix interface and preferred to glide along the interface rather than shear the precipitate and similar observations were provided by molecular dynamics simulations [47]. It should be noticed that first principles calculations of the generalized stacking fault energy have shown that high shear stresses are necessary to promote dislocation slip in these intermetallic phases [49] and, thus, several dislocations have to pile-up before the Orowan loop can penetrate and collapse within the precipitate. Moreover, precipitate shearing by basal dislocations in Mg alloys is favored because the Mg basal plane is often parallel to one crystallographic plane of the precipitate which tend to form coherent interfaces along the basal plane, which is the most closely packed. These crystallographic orientations corresponds to $(0001)_{Mg}\|(2\bar{1}\bar{1}0)_{\beta_1'}$ in Mg-Zn and Mg-Nd alloys and to $(0001)_{Mg}\|(110)_{Mg_{17}Al_{12}}$ in Mg-Al alloys and lead to the shearing of very thick precipitates (as the one shown in Fig. 9c) by basal dislocations. Nevertheless, precipitate shearing in Mg alloys by either prismatic or pyramidal dislocations has not been observed because of the lack of crystallographic continuity between the corresponding slip planes in the matrix and in the precipitate [45].

The first consequence of precipitate shearing is the limited strain hardening of Mg alloys, as opposed to the strong hardening associated by the formation of dislocation pile-ups in front of impenetrable precipitates in the absence of precipitate shearing. In addition, precipitate shearing facilitates the localization of deformation along slip bands (Figs. 6b and 9a) and both mechanisms were responsible for the constant value of the CRSS for strains > 2%.

From the qualitative viewpoint, the mechanisms of dislocation/precipitate interaction did not change between 23 ºC and 100 ºC. Nevertheless, it is obvious that precipitate shearing was easier at high temperature, as demonstrated for the higher number of slip bands and the fragmentation of the precipitates within the slip bands as a result of multiple shearing along parallel planes.

### 4.2. Strengthening mechanisms

The initial CRSS to move dislocations in the basal plane can be computed from the addition of three different contributions.

$$\tau_{CRSS} = \tau_0 + \tau_{SS} + \tau_P \qquad (2)$$

where $\tau_0$ is the CRSS to move dislocations in the basal plane of pure Mg, $\tau_{SS}$ is the solid solution contribution and $\tau_P$ is due to the interaction of the dislocations with the precipitates.

$\tau_0$ is known to be very low for basal slip in Mg (≈ 0.5 MPa) [7,8] while $\tau_{SS}$ was determined experimentally in Mg-Zn alloys from mechanical tests in single crystals [37] and micropillar compression tests [28] and it was found to be depended on the Zn content according to

$$\tau_{SS} = Kc^{2/3} \qquad (3)$$

where K = 3.48, $c$ stands for the atomic concentration of Zn (in %) and $\tau_{SS}$ is expressed in MPa [28]. The initial Zn content in solid solution in the alloy was 1.5 at. % Zn (equivalent to 4 wt. %) which was reduced to approximately 0.8 at. % of Zn after the aging treatments, taking into account the amount of Zn necessary to form the precipitates. Thus, the contribution to the initial

CRSS of the Zn in solid solution was ≈ 3 MPa for both aging conditions according to eq. (3) and it was assumed to be independent of the temperature in the range 23ºC to 100ºC.

The experimental observations presented above showed evidence of dislocations bowing between precipitates and precipitate shearing. The strengthening contribution due to the formation of Orowan loops around the rod-shape precipitates can be quantified according to [43]

$$\tau_{Orowan} = \frac{Gb}{2\pi\lambda\sqrt{1-\nu}} \ln\frac{d}{r_0} \qquad (4)$$

where $G$ and $\nu$ are the shear modulus and Poisson's ratio of Mg, $b$ the magnitude of the Burgers vector, $\lambda$ the average spacing between precipitates, $d$ the average diameter of the precipitates and $r_0$ the core radius of dislocations. Assuming $r_0 = b$ and taking into account that $G$ = 16.6 GPa, $b$ = 3.3 Å, $\nu$ = 0.33, and the precipitate spacing and diameter given in section 3.1, $\tau_{Orowan}$ = 44.6 ± 5.5 MPa and 27.2 ± 3.2 MPa for the alloys aged at 149ºC and 204ºC, respectively.

Adding the $\tau_0$ and $\tau_{SS}$ contributions, the predicted values of the initial CRSS given by Eq. (2) were 48.1 ± 5.5 MPa and 30.7± 3.2 MPa for the alloys aged at 149 ºC and 204 ºC, respectively. These values were close to the initial CRSS values measured at 23 ºC in both aging conditions (Table 1) and they are compatible with the experimental observations of dislocations bowing between the precipitates. Further deformation led to formation of dislocation pile-ups in front of the precipitates, and the CRSS increased up to the plateau observed in all the stress-strain curves in Fig. 5. The accumulation of dislocations led to the shearing of the precipitates by the basal dislocations, in agreement with the atomistic simulations [47].

The CRSS to shear the precipitates depends on a number of factors, which include chemical or interface strengthening, modulus mismatch strengthening, stacking fault energy mismatch strengthening, coherency strain strengthening and order strengthening [50]. Unfortunately, the models available to predict the contribution of each mechanism are not very reliable or depend on parameters that are difficult to estimate [45-46]. Nevertheless, the contribution of chemical strengthening due to creation of more interface area when the dislocation shears the precipitate is known to be very small for all systems and can be neglected. Similarly, the mismatch in the shear modulus along the slip plane of Mg (16.6 GPa) and MgZn$_2$ (25 GPa [47]) is small. Regarding coherency strains, they may have a large impact which affects dislocation motion whether they shear or bow around the precipitates [50], but they have never been found in the interaction of non-basal <a> and <a+c> dislocations with precipitates [45] and it seems reasonable to assume that they are small. Finally, no superlattice correspondence was found between the (0001) basal plane of Mg and the $(2\bar{1}\bar{1}0)$ prismatic plane of the $\beta_1'$ precipitates and thus order strengthening should not be relevant in this case. Thus, stacking fault energy strengthening should be the most relevant mechanism in this case. This contribution is proportional to the differences in the stacking fault energy between the matrix and the precipitate, which decreases as the temperature increases [51]. This reduction in the differences in stacking fault energies with temperature would be responsible for the lower values of the CRSS at 100 ºC, as compared with those measured at 23 ºC. Further evidence showing that precipitate shearing was easier at 100 ºC (as compared with 23ºC) is found in Figs. 9a and 9d. They show that more slip bands were formed along the micropillar and that the precipitates were transformed into globular particles within the slip bands.

## 5. Conclusions

The effect of $\beta_1'$ MgZn$_2$ precipitates on the critical resolved shear stress for basal slip and on the dislocation/precipitate interactions was analyzed in an Mg-4 wt.% Zn alloy at 23 ºC and 100 ºC by means of micropillar compression tests. The main conclusions of this investigation are the following:

- It was found that the initial CRSS and the CRSS at 4% strain were independent of the micropillar dimensions for micropillars with a cross-section equal to or larger than 5 × 5 μm$^2$ because these dimensions were much larger than the precipitate spacing. Thus, the CRSS for basal slip in the bulk alloy was obtained from these tests as a function of the precipitate spacing and temperature.
- Transmission electron microscopy showed that the interactions between basal dislocations and $\beta_1'$ precipitates involved a mixture of dislocation bowing around the precipitates and precipitate shearing. The latter did not take place along well-defined crystallographic planes but involved the deformation of the precipitate along a region encompassing several atomic planes.
- As a result of shearing, deformation was localized in slip bands leading to constant CRSS at strains larger than 2%. Both the initial critical resolved shear stress and the plateau stress decreased significantly (by 34% to 35%) between 23 ºC and 100 ºC. The initial CRSS at 23 ºC was in good agreement with the predictions of the Orowan model assuming that the dislocations bow around the precipitates but the reduction in the CRSS with temperature was associated to the increase in the number of slip bands and in the fragmentation of the precipitates due to successive shearing by the dislocations.

**Acknowledgements**

This investigation was supported by the European Research Council (ERC) under the European Union's Horizon 2020 research and innovation programme (Advanced Grant VIRMETAL, grant agreement No. 669141. R. Alizadeh also acknowledges the support from the Spanish Ministry of Science through the Juan de la Cierva program (FJCI-2016-29660). The authors acknowledge the support of Dr. Jingya Wang, Dr. Miguel Monclus and Dr. Miguel Castillo for their helps to carry out the high temperature micropillar compression tests and the TEM observations.

**References**

[1] B.L. Mordike, T. Ebert, Magnesium Properties - applications - potential, Mater. Sci. Eng. A. 302 (2001) 37–45. doi:10.1016/S0921-5093(00)01351-4.

[2] M.P. Staiger, A.M. Pietak, J. Huadmai, G. Dias, Magnesium and its alloys as orthopedic biomaterials: A review, Biomaterials. 27 (2006) 1728–1734. doi:10.1016/j.biomaterials.2005.10.003.

[3] M.K. Kulekci, Magnesium and its alloys applications in automotive industry, Int. J. Adv. Manuf. Technol. 39 (2008) 851–865. doi:10.1007/s00170-007-1279-2.

[4] R. Alizadeh, R. Mahmudi, Evaluating high-temperature mechanical behavior of cast Mg-4Zn-xSb magnesium alloys by shear punch testing, Mater. Sci. Eng. A. 527 (2010) 3975–3983. doi:10.1016/j.msea.2010.03.007.

[5] R. Alizadeh, R. Mahmudi, Effect of Sb additions on the microstructural stability and mechanical properties of cast Mg-4Zn alloy, Mater. Sci. Eng. A. 527 (2010) 5312–5317. doi:10.1016/j.msea.2010.05.029.

[6] E.C. Burke, W.R. Hibbard, Plastic deformation of magnesium single crystals, Trans. Met. Soc. AIME. 194 (1952) 295–303.

[7] H. Conrad, W.D. Robertson, Effect of temperature on the flow stress and strain-hardening coefficient of magnesium single crystals, Jom. 9 (1957) 503–512. doi:10.1007/BF03397908.

[8] H. Yoshinaga, R. Horiuchi, Deformation mechanisms in magnesium single crystals compressed in the direction parallel to hexagonal axis, Trans. JIM. 4 (1963) 1–8. doi:10.2320/matertrans1960.4.1.

[9] R.E. Reed-Hill, W.D. Robertson, Deformation of magnesium single crystals by nonbasal slip, Jom. 9 (1957) 496–502. doi:10.1007/BF03397907.

[10] T. Obara, H. Yoshinga, S. Morozumi, {11$\bar{2}$2} ⟨1123⟩ Slip system in magnesium, Acta Metall. 21 (1973) 845–853. doi:10.1016/0001-6160(73)90141-7.

[11] C.M. Cepeda-Jiménez, J.M. Molina-Aldareguia, M.T. Pérez-Prado, Effect of grain size on slip activity in pure magnesium polycrystals, Acta Mater. 84 (2015) 443–456. doi:10.1016/j.actamat.2014.10.001.

[12] S. Sandlöbes, M. Friák, J. Neugebauer, D. Raabe, Basal and non-basal dislocation slip in Mg–Y, Mater. Sci. Eng. A. 576 (2013) 61–68. doi:10.1016/j.msea.2013.03.006.

[13] A. Akhtar, E. Teghtsoonian, Substitutional solution hardening of magnesium single crystals, Philos. Mag. 25 (1972) 897–916. doi:10.1080/14786437208229311.

[14] S. Miura, S. Imagawa, T. Toyoda, K. Ohkubo, T. Mohri, Effect of Rare-Earth Elements Y and Dy on the Deformation Behavior of Mg Alloy Single Crystals, Mater. Trans. 49 (2008) 952–956. doi:10.2320/matertrans.mc2007109.

[15] A. Akhtar, E. Teghtsoonian, Solid Solution Strengthening of Magnesium Single Crystals—I Alloying Behaviour in Basal Slip, Acta Metall. 17 (1969) 1339–1349. doi:10.1016/0001-6160(69)90151-5.

[16] Y. Liu, N. Li, M. Arul Kumar, S. Pathak, J. Wang, R.J. McCabe, N.A. Mara, C.N. Tom?, Experimentally quantifying critical stresses associated with basal slip and twinning in magnesium using micropillars, Acta Mater. (2017). doi:10.1016/j.actamat.2017.06.008.

[17] Y. Liu, N. Li, M. Arul Kumar, S. Pathak, J. Wang, R.J. McCabe, N.A. Mara, C.N. Tom?, Experimentally quantifying critical stresses associated with basal slip and twinning in magnesium using micropillars, Acta Mater. 135 (2017) 411–421. doi:10.1016/j.actamat.2017.06.008.

[18] J. Bočan, S. Tsurekawa, A. Jäger, Fabrication and in situ compression testing of Mg micropillars with a nontrivial cross section: Influence of micropillar geometry on mechanical properties, Mater. Sci. Eng. A. 687 (2017) 337–342. doi:10.1016/j.msea.2017.01.089.

[19] K.E. Prasad, K. Rajesh, U. Ramamurty, Micropillar and macropillar compression responses of magnesium single crystals oriented for single slip or extension twinning, Acta Mater. 65 (2014) 316–325. doi:10.1016/j.actamat.2013.10.073.

[20] C.M. Byer, K.T. Ramesh, Effects of the initial dislocation density on size effects in single-crystal magnesium, Acta Mater. 61 (2013) 3808–3818. doi:10.1016/j.actamat.2013.03.019.

[21] E. Lilleodden, Microcompression study of Mg (0 0 0 1) single crystal, Scr. Mater. 62 (2010) 532–535. doi:10.1016/j.scriptamat.2009.12.048.

[22] C.M. Byer, B. Li, B. Cao, K.T. Ramesh, Microcompression of single-crystal magnesium, Scr. Mater. 62 (2010) 536–539. doi:10.1016/j.scriptamat.2009.12.017.

[23] J.J. Williams, J.L. Walters, M.Y. Wang, N. Chawla, A. Rohatgi, Extracting constitutive stress-strain behavior of microscopic phases by micropillar compression, Jom. 65 (2013) 226–233. doi:10.1007/s11837-012-0516-9.

[24] J. Wang, N. Stanford, Investigation of precipitate hardening of slip and twinning in Mg5%Zn by micropillar compression, Acta Mater. 100 (2015) 53–63.

doi:10.1016/j.actamat.2015.08.012.

[25] W.S. Chuang, C.H. Hsieh, J.C. Huang, P.H. Lin, K. Takagi, Y. Mine, K. Takashima, Relation between sample size and deformation mechanism in Mg-Zn-Y 18R-LPSO single crystals, Intermetallics. 91 (2017) 110–119. doi:10.1016/j.intermet.2017.08.009.

[26] J. Ye, R.K. Mishra, A.K. Sachdev, A.M. Minor, In situ TEM compression testing of Mg and Mg-0.2 wt.% Ce single crystals, Scr. Mater. 64 (2011) 292–295. doi:10.1016/j.scriptamat.2010.09.047.

[27] M. Pozuelo, Y.W. Chang, J.M. Yang, In-situ microcompression study of nanostructured Mg alloy micropillars, Mater. Lett. 108 (2013) 320–323. doi:10.1016/j.matlet.2013.05.039.

[28] J.Y. Wang, N. Li, R. Alizadeh, M.A. Monclús, Y.W. Cui, J.M. Molina-Aldareguía, J. LLorca, Effect of solute content and temperature on the deformation mechanisms and critical resolved shear stress in Mg-Al and Mg-Zn alloys, Acta Mater. 170 (2019) 155–165. doi:10.1016/j.actamat.2019.03.027.

[29] J.R. Greer, W.C. Oliver, W.D. Nix, Size dependence of mechanical properties of gold at the micron scale in the absence of strain gradients, Acta Mater. 53 (2005) 1821–1830. doi:10.1016/j.actamat.2004.12.031.

[30] J.R. Greer, J.T.M. De Hosson, Plasticity in small-sized metallic systems: Intrinsic versus extrinsic size effect, Prog. Mater. Sci. 56 (2011) 654–724. doi:10.1016/j.pmatsci.2011.01.005.

[31] A.H.W. Ngan, An explanation for the power-law scaling of size effect on strength in micro-specimens, Scr. Mater. 65 (2011) 978–981. doi:10.1016/j.scriptamat.2011.08.027.

[32] R. Gu, A.H.W. Ngan, Effects of pre-straining and coating on plastic deformation of aluminum micropillars, Acta Mater. 60 (2012) 6102–6111. doi:10.1016/j.actamat.2012.07.048.

[33] A. Cruzado, B. Gan, M. Jiménez, D. Barba, K. Ostolaza, A. Linaza, J.M. Molina-Aldareguia, J. Llorca, J. Segurado, Multiscale modeling of the mechanical behavior of IN718 superalloy based on micropillar compression and computational homogenization, Acta Mater. 98 (2015) 242–253. doi:10.1016/j.actamat.2015.07.006.

[34] A. Singh, J.M. Rosalie, H. Somekawa, T. Mukai, The structure of β1 precipitates in Mg-Zn-Y alloys, Philos. Mag. Lett. 90 (2010) 641–651. doi:10.1080/09500839.2010.490049.

[35] J.-F. Nie, Precipitation and Hardening in Magnesium Alloys, Metall. Mater. Trans. A. 43 (2012) 3891–3939. doi:10.1007/s11661-012-1217-2.

[36] J.F. Nie, Physical Metallurgy of Light Alloys, in: Phys. Metall., 2014: pp. 2009–2156.

[37] D.S. Chun, G. Byrne, Precipitate Strengthening Mechanisms in Magnesium Zinc Alloy Single Crystals, J. Mater. Sci. 4 (1969) 861–872. doi:10.1007/BF00549777.

[38] J.B. Clark, Transmission electron microscopy study of age hardening in a Mg-5 wt.% Zn alloy, Acta Metall. 13 (1965) 1281–1289.

[39] D. Kiener, C. Motz, G. Dehm, Micro-compression testing: A critical discussion of

experimental constraints, Mater. Sci. Eng. A. 505 (2009) 79–87. doi:10.1016/j.msea.2009.01.005.

[40] H. Zhang, B.E. Schuster, Q. Wei, K.T. Ramesh, The design of accurate micro-compression experiments, Scr. Mater. 54 (2006) 181–186. doi:10.1016/j.scriptamat.2005.06.043.

[41] J. Hütsch, E.T. Lilleodden, The influence of focused-ion beam preparation technique on microcompression investigations: Lathe vs. annular milling, Scr. Mater. 77 (2014) 49–51. doi:10.1016/j.scriptamat.2014.01.016.

[42] I.N. Sneddon, The relation between load and penetration in the axisymmetric Boussinesq problem for a punch of arbitrary profile, Int. J. Eng. Sci. 3 (1965) 47–57. doi:10.1016/0020-7225(65)90019-4.

[43] J.F. Nie, Effects of precipitate shape and orientation on dispersion strengthening in magnesium alloys, Scr. Mater. 48 (2003) 1009–1015. doi:10.1016/S1359-6462(02)00497-9.

[44] B. Zhou, L. Wang, G. Zhu, J.I.E. Wang, Understanding the Strengthening Effect of b 1 Precipitates in Mg-Nd Using In Situ Synchrotron X-ray Diffraction, JOM. (2018). doi:10.1007/s11837-018-2972-3.

[45] J.J. Bhattacharyya, F. Wang, N. Stanford, S.R. Agnew, Slip mode dependency of dislocation shearing and looping of precipitates in Mg alloy WE43, Acta Mater. 146 (2018) 55-62. doi:10.1016/j.actamat.2017.12.043.

[46] Z. Huang, C. Yang, L. Qi, J.E. Allison, A. Misra, Dislocation pile-ups at $\beta_1$ precipitate interfaces in Mg-rare earth (RE) alloys, Mater. Sci. Eng. A. 742 (2019) 278–286. doi:10.1016/j.msea.2018.10.104.

[47] G. Esteban-Manzanares, A. Ma, I. Papadimitriou, E. Martínez, J. LLorca. Basal dislocation/precipitate interactions in Mg-Al alloys: an atomistic investigation. Modelling and Simulation in Materials Science and Engineering, 27, 075003, 2019. doi: 10.1088/1361-651X/ab2de0

[48] Z. Huang, J.E. Allison, A. Misra, Interaction of Glide Dislocations with Extended Precipitates in Mg-Nd alloys, Sci. Rep. 8 (2018) 1–12. doi:10.1038/s41598-018-20629-1.

[49] W. Xiao, X. Zhang, W.T. Geng, G. Lu, Atomistic study of plastic deformation in Mg-Al alloys, Mater. Sci. Eng. A. 586 (2013) 245–252. doi:10.1016/j.msea.2013.07.093.

[50] R. Santos-Güemes, G. Esteban-Manzanares, I. Papadimitriou, J. Segurado, L. Capolungo, J. LLorca, Discrete dislocation dynamics simulations of dislocation-θ′ precipitate interaction in Al-Cu alloys, J. Mech. Phys. Solids. 118 (2018) 228–244. doi:10.1016/j.jmps.2018.05.015.

[51] E. Nembach, Particle Strengthening of Metals and Alloys, Jon Wiley & Sons, New York, 1997.

**SUPPLEMENTARY MATERIAL**

**Fig. S1**. STEM tomography of the material aged at 149 °C before deformation, showing the rod-shaped $\beta_1'$ precipitates which grew parallel to the *c*-axis of the Mg matrix.

**Fig. S2**. STEM tomography of the material aged at 149 °C, showing the morphology of the $\beta_1'$ precipitates after room-temperature deformation. This tomography was obtained from an area between slip bands (1) and (2) in Fig 7a.

**Fig. S3**. STEM tomography of the material aged at 149 °C, showing the morphology of the $\beta_1'$ precipitates after deformation at 100 °C. This tomography was obtained from an area between slip bands (2) and (4) in Fig 9a.

**Fig. S4**. STEM tomography of the material aged at 149 °C, showing the morphology of the $\beta_1'$ precipitates after deformation at 100 °C. This tomography was obtained from an area between slip bands (1) and (2) in Fig 9d.